\documentclass[showpacs,amsmath,amssymb,prd,floatfix,twocolumn,superscriptaddress,nofootinbib]{revtex4}
\usepackage{graphicx}
\usepackage{epsfig}
\usepackage{bm}
\usepackage{amsfonts}
\usepackage[draft=false]{hyperref}
\usepackage{threeparttable}
\usepackage{latexsym}
\newcommand{\s}{\phi}
\newcommand{\la}{\lambda}

\newcommand{\be}{\begin{equation}}
\newcommand{\ee}{\end{equation}}

\begin{document}

\title{Phantom dark energy with varying-mass dark matter particles: acceleration and cosmic coincidence problem}

\author{Genly Leon}
\email{genly@uclv.edu.cu} \affiliation{Department of Mathematics,
Universidad Central de Las Villas, Santa Clara CP 54830, Cuba}

\author{Emmanuel N. Saridakis }
\email{msaridak@phys.uoa.gr} \affiliation{Department of Physics,
University of Athens, GR-15771 Athens, Greece}

\begin{abstract}
We investigate several varying-mass dark-matter particle models in
the framework of phantom cosmology. We examine whether there exist
late-time cosmological solutions, corresponding to an accelerating
universe and possessing dark energy and dark matter densities of
the same order. Imposing exponential or power-law potentials and
exponential or power-law mass dependence, we conclude that the
coincidence problem cannot be solved or even alleviated. Thus, if
dark energy is attributed to the phantom paradigm, varying-mass
dark matter models cannot fulfill the basic requirement that led
to their construction.
 \end{abstract}


 \maketitle

\section{Introduction}

Recent cosmological observations support that the universe is
experiencing an accelerated expansion, and that the transition to
the accelerated phase realized in the recent cosmological past
\cite{observ}. In order to explain this unexpected behavior, one
can modify the theory of gravity \cite{ordishov}, or introduce the
concept of dark energy which provides the acceleration mechanism.
The most explored dynamical dark energy models of the literature
consider a canonical scalar field (quintessence) \cite{quint}, a
phantom field, that is a scalar field with a negative sign of the
kinetic term \cite{phant}, or the combination of quintessence and
phantom in a unified model named quintom \cite{quintom}.

The dynamical nature of dark energy introduces a new cosmological
problem, namely why are the densities of vacuum energy and dark
matter nearly equal today although they scale independently during
the expansion history. The elaboration of this ``coincidence''
problem led to the consideration of generalized versions of the
aforementioned scenarios with the inclusion of a coupling between
dark energy and dark matter. Thus, various forms of
``interacting'' dark energy models
\cite{interacting,interacting2,Guo:2004xx,ChenSaridakis} have been
constructed in order to fulfil the observational requirements. In
the case of interacting quintessence one can find accelerated
attractors which moreover give dark matter and dark energy density
parameters of the same order,  thus solving the coincidence
problem \cite{Wetterich:1994bg,Wetterich:1994bg2}, but paying the
price of introducing new problems such is the justification of a
non-trivial, almost tuned, sequence of cosmological epochs
\cite{Amendola:2006qi}. In interacting phantom models
\cite{Guo:2004vg,Guo:2004xx,ChenSaridakis}, the existing
literature remains in some special coupling forms which suggest
that the coincidence problem might be alleviated
\cite{Guo:2004xx,ChenSaridakis}.

An equivalent approach is to assume that dark energy and dark
matter sectors interact in such a way that the dark matter
particles acquire a varying mass, dependent on the scalar field
which reproduces dark energy \cite{Anderson:1997un}. This
consideration allows for a better theoretical justification, since
a scalar-field-dependent varying-mass can arise from string or
scalar-tensor theories \cite{Damour:1990tw}. Indeed, in such
higher dimensional frameworks one can formulate both the
appearance of the scalar field (which is related to the dilaton
and moduli fields) and its effect on  matter particle masses
(determined by string dynamics, supersymmetry breaking, and the
compactification mechanism) \cite{Casas:1991ky}. In quintessence
scenario, such varying-mass dark matter models have been explored
in cases of linear
\cite{Anderson:1997un,interacting2,Casas:1991ky,quirosvamps},
power-law \cite{Zhang:2005rg} or exponential
\cite{Amendola:1999er,Amendola:2001rc,Comelli2003} scalar-field
dependence. The exponential case is the most interesting since,
apart from solving the coincidence problem, it allows for stable
scaling behavior, that is for a large class of initial conditions
the cosmological evolution converges to a common solution at late
times \cite{Amendola:1999er,Comelli2003}.

In the present work we are interested in investigating
varying-mass dark matter models in scenarios where dark energy is
attributed to a phantom field. Although such a framework could
lead to instabilities at the quantum level \cite{Cline:2003gs},
there have been serious attempts in overcoming these difficulties
and construct a phantom theory consistent with the basic
requirements of quantum field theory, with the phantom fields
arising as an effective description \cite{quantumphantom0}.
Performing a complete phase-space analysis using various forms of
mass-dependence and scalar-field potentials, we examine whether
there exist stable late-time accelerating solutions which moreover
solve the coincidence problem. As we will show, the coincidence
problem cannot be solved in any of the investigated models.

The plan of the work is as follows: In section \ref{phantcosm} we
construct varying-mass dark matter models in the framework of
phantom cosmological scenario and we present the formalism for the
transformation into an autonomous dynamical system. In section
\ref{models} we perform the phase-space stability analysis for
four  different models, using various mass-dependence forms and
phantom potentials, and in section \ref{cosmimpl} we discuss the
corresponding cosmological implications. Finally, in section
\ref{conclusions} we summarize the obtained results.

\section{Varying-mass dark matter particles in the framework of phantom cosmology}
\label{phantcosm}

Let us construct a cosmological model where dark energy is
attributed to a phantom field, in which the dark matter particles
have a varying mass depending on this field. Throughout the work
we consider a flat Robertson-Walker metric:
\begin{equation}\label{metric}
ds^{2}=dt^{2}-a^{2}(t)d\bf{x}^2,
\end{equation}
with $a$ the scale factor and  $t$ the comoving time.

In the phantom cosmological paradigm the energy density and
pressure of the phantom scalar field $\phi$ are:
\begin{eqnarray}\label{rhophi}
 \rho_{\phi}&=& -\frac{1}{2}\dot{\phi}^{2} + V(\phi)\\
 \label{pphi}
 p_{\phi}&=& - \frac{1}{2}\dot{\phi}^{2} - V(\phi),
\end{eqnarray}
where $V(\phi)$ is the phantom potential and the dot denotes
differentiation with respect to comoving time. In such a scenario,
the dark energy is attributed to the phantom field, and its
equation of state is given by
\begin{equation}
w_{DE}\equiv w_{\phi}=\frac{p_\phi}{\rho_\phi}.
\end{equation}

As was mentioned in the introduction,
 in varying-mass dark matter models the central assumption is that the
dark-matter particles have a $\phi$-dependent mass $M_{DM}(\phi)$,
while dark matter is considered as dust. Thus, for the dark matter
energy density we have the standard definition
\begin{equation}\label{rhodm}
\rho_{DM}=M_{DM}(\phi)\,n_{DM},
\end{equation}
where $n_{DM}$ is the number density of the dark-matter particles.
As usual, in the case of FRW geometry, it is determined by the
equation
\begin{equation}\label{ndmm}
\dot{n}_{DM}+3Hn_{DM}=0,
\end{equation}
 with $H$ the Hubble parameter. Therefore, differentiating
 (\ref{rhodm}) and using  (\ref{ndmm}) we obtain the evolution
 equation for $\rho_{DM}$, namely:
\begin{equation}\label{rhodmeom}
\dot{\rho}_{DM}+3H\rho_{DM}=\frac{1}{M_{DM}(\phi)}\frac{dM_{DM}(\phi)}{d\phi}\,\dot{\phi}\,\rho_{DM}.
\end{equation}
Obviously, in a case of $\phi$-independent dark-matter particle
mass, we re-obtain the usual evolution equation
$\dot{\rho}_{DM}+3H\rho_{DM}=0$. Therefore, we observe that the
$\phi$-dependent mass reveals the interaction between dark matter
and dark energy (that is the phantom field) sectors that lies
behind it.

Since general covariance leads to total energy conservation, we
deduce that the evolution equation for the phantom energy density
will be:
\begin{equation}\label{rhophieom}
\dot{\rho}_\phi+3H(\rho_\phi+p_\phi)=-\frac{1}{M_{DM}(\phi)}\frac{dM_{DM}(\phi)}{d\phi}\,\dot{\phi}\,\rho_{DM}.
\end{equation}
Thus, $\frac{dM_{DM}(\phi)}{d\phi}\,\dot{\phi}<0$ corresponds to
energy transfer from dark matter to dark energy, while
$\frac{dM_{DM}(\phi)}{d\phi}\,\dot{\phi}>0$ corresponds to dark
energy transformation into dark matter.

Equivalently, using the definitions (\ref{rhophi}) and
(\ref{pphi}), the phantom evolution equation can be written in
field terms as:
\begin{equation}\label{phiddot}
\ddot{\phi}+3H\dot{\phi}-\frac{\partial
V(\phi)}{\partial\phi}=\frac{1}{M_{DM}(\phi)}\frac{dM_{DM}(\phi)}{d\phi}\,\rho_{DM}.
\end{equation}
Finally, the system of equations closes by considering the
Friedmann equations:
\begin{equation}\label{FR1}
H^{2}=\frac{\kappa^{2}}{3}(\rho_{\phi}+\rho_{DM}),
\end{equation}
\begin{equation}\label{FR2}
\dot{H}=-\frac{\kappa^2}{2}\Big(\rho_{\phi}+p_{\phi}+\rho_{DM}\Big),
\end{equation}
where we have set $\kappa^2\equiv 8\pi G$. Although we could
straightforwardly include baryonic matter and radiation in the
model, for simplicity reasons we neglect them.

Alternatively, one could construct the equivalent uncoupled model
described by:
\begin{eqnarray}
\dot{\rho}_{DM}+3H(1+w_{DM,eff})\rho_{DM}=0\\
\dot{\rho}_\phi+3H(1+w_{\phi,eff})\rho_\phi=0,
\end{eqnarray}
where
\begin{eqnarray}
&&w_{DM,eff}=-\frac{1}{M_{DM}(\phi)}\frac{dM_{DM}(\phi)}{d\phi}\,\frac{\dot{\phi}}{3H}\\
&&w_{\phi,eff}=w_\phi
+\frac{1}{M_{DM}(\phi)}\frac{dM_{DM}(\phi)}{d\phi}\,\frac{\dot{\phi}}{3H}\,\frac{\rho_{DM}}{\rho_\phi}.\
\ \ \
\end{eqnarray}
However, it is more convenient to introduce the ``total''  energy
density $\rho_{tot}\equiv\rho_{DM}+\rho_\phi$, obtaining:
\begin{equation}
\label{rhot}
 \dot{\rho}_{tot}+3 H(1+w_{tot})\rho_{tot}=0,
\end{equation}
with
\begin{equation}
w_{tot}=\frac{p_\phi}{\rho_\phi+\rho_{DM}}=w_\phi\Omega_\phi,
\end{equation}
where
$\Omega_\phi\equiv\frac{\rho_\phi}{\rho_{tot}}\equiv\Omega_{DE}$.
Obviously, since $\rho_{tot}=3H^2/\kappa^2$, (\ref{rhot}) leads to
a scale factor evolution of the form $a(t)\propto
t^{2/(3(1+w_{tot}))}$, in the constant $w_{tot}$ case. However, at
the late-time stationary solutions that we are studying in the
present work, $w_{tot}$ has reached to a constant value and thus
the above behavior is valid. Therefore, we conclude that in such
stationary solutions the condition for acceleration is just
$w_{tot}<-1/3$.

In order to perform the phase-space and stability analysis of the
phantom model at hand, we have to transform the aforementioned
dynamical system into its autonomous form
\cite{expon,Copeland:1997et}. This will be achieved by introducing
the auxiliary variables:
\begin{eqnarray}
&&x=\frac{\kappa\dot{\phi}}{\sqrt{6}H},\nonumber\\
&&y=\frac{\kappa\sqrt{V(\phi)}}{\sqrt{3}H}, \nonumber\\
&&z=\frac{\sqrt{6}}{\kappa\phi} \label{auxilliary}
\end{eqnarray}
together with $M=\ln a$. Thus, it is easy to see that for every
quantity $F$ we acquire $\dot{F}=H\frac{dF}{dM}$.
 Using these
variables we obtain:
\begin{equation}
 \Omega_{\phi}\equiv\frac{\kappa^{2}\rho_{\phi}}{3H^{2}}=-x^2+y^2,
 \label{Omegas}
\end{equation}
\begin{equation}\label{wss}
w_{\phi}=\frac{-x^2-y^2}{-x^2+y^2},
\end{equation}
and
\begin{equation}\label{wtot}
w_{tot}=-x^2-y^2.
\end{equation}
We mention that relations (\ref{wss}) and (\ref{wtot}) are always
valid, that is independently of the specific state of the system
(they are valid in the whole phase-space and not only at the
critical points). Finally, note that in the case of complete dark
energy domination, that is $\rho_{DM}\rightarrow0$ and
$\Omega_\phi\rightarrow1$, we acquire $w_{tot}\approx
w_\phi\leq-1$, as expected to happen in phantom-dominated
cosmology.

The next step is the introduction of a specific ansatz for the
phantom potential $V(\phi)$, and a specific ansatz for the
dark-matter particle mass function $M_{DM}(\phi)$.
 In this case the equations of motion
(\ref{rhodmeom}), (\ref{phiddot}), (\ref{FR1}) and (\ref{FR2}) can
be transformed into an autonomous system containing the variables
$x$ and $y$ and perhaps $z$ ($z$ is present only for some such
ansatzes) and their derivatives with respect to $M=\ln a$.

Having transformed the cosmological system into its autonomous
form:
\begin{equation}\label{eomscol}
\textbf{X}'=\textbf{f(X)},
\end{equation}
where $\textbf{X}$ is the column vector constituted by the
auxiliary variables, \textbf{f(X)} the corresponding  column
vector of the autonomous equations, and prime denotes derivative
with respect to $M=\ln a$, we extract its critical points
$\bf{X_c}$  satisfying $\bf{X}'=0$. Then, in order to determine
the stability properties of these critical points, we expand
(\ref{eomscol}) around  $\bf{X_c}$, setting
$\bf{X}=\bf{X_c}+\bf{U}$ with $\textbf{U}$ the perturbations of
the variables considered as a column vector. Thus, for each
critical point we expand the equations for the perturbations up to
the first order as:
\begin{eqnarray}
\label{perturbation} \textbf{U}'={\bf{Q}}\cdot \textbf{U},
\end{eqnarray}
where the matrix ${\bf {Q}}$ contains the coefficients of the
perturbation equations.
 Thus,
for each critical point, the eigenvalues of ${\bf {Q}}$ determine
its type and stability.

\section{Phase-space analysis} \label{models}

In the previous section we constructed a cosmological scenario
where the dark matter particles have a varying mass, depending on
the phantom field. Additionally, we presented the formalism for
its transformation into an autonomous dynamical system, suitable
for a stability analysis. In this section we introduce specific
forms for $V(\phi)$ and $M_{DM}(\phi)$, and we perform a complete
phase-space analysis.

For the scalar field potential we consider two well studied cases
of the literature, namely the exponential
\cite{Amendola:1999er,Comelli2003}:
\begin{equation}\label{exppot}
V(\phi)=V_0e^{-\kappa\lambda_1\phi}
\end{equation}
and the power-law one \cite{Zhang:2005rg,Kneller:2003xg}:
\begin{equation}\label{powerpot}
V(\phi)=V_0\phi^{-\lambda_2}.
\end{equation}
For the  dark matter particle mass we consider two possible cases,
namely an exponential dependence
\cite{Amendola:1999er,Amendola:2001rc,Comelli2003}:
\begin{equation}\label{expmass}
M_{DM}(\phi)=M_0e^{-\kappa\mu_1\phi}
\end{equation}
and the power-law one \cite{Zhang:2005rg}:
\begin{equation}\label{powermass}
M_{DM}(\phi)=M_0\phi^{-\mu_2}.
\end{equation}
Therefore, in the following we consider four different models,
arising from the aforementioned combinations.

\subsection{Model 1: Exponential potential and exponentially-dependent dark-matter particle mass}

Inserting the auxiliary variables (\ref{auxilliary}) into the
equations of motion (\ref{rhodmeom}), (\ref{phiddot}), (\ref{FR1})
and (\ref{FR2}), we result in the following autonomous system:
\begin{eqnarray}
x'=-
3x+\frac{3}{2}x (1-x^2-y^2)-\sqrt{\frac{3}{2}}\la_1\, y^2-\ \ \ \ \ \ \   \ \nonumber\\
-\sqrt{\frac{3}{2}}\mu_1(1+x^2-y^2)\nonumber\\
y'=\frac{3}{2}y (1-x^2-y^2)- \sqrt{\frac{3}{2}}\la_1\, x y
\label{autonomous1}.\ \ \ \ \ \ \ \ \ \ \ \ \ \   \ \   \ \ \ \
\end{eqnarray}
Note that in this case, the auxiliary variable $z$ is not needed.

The critical points $(x_c,y_c)$ of the autonomous system
(\ref{autonomous1}) are obtained by setting the left hand sides of
the equations to zero. The real and physically meaningful (that is
corresponding to $y>0$ and $0\leq\Omega_\phi\leq1$) of them are:
\begin{eqnarray}
\left(x_{c1}=-\frac{\la_1}{\sqrt{6}},\
y_{c1}=\sqrt{1+\frac{\la_1^2}{6}}\right), \nonumber
\end{eqnarray}
\begin{equation}\left(x_{c2}=\frac{\sqrt{\frac32}}{\la_1-\mu_1},\
y_{c2}=\frac{\sqrt{-\frac32-\mu_1\left(\la_1-\mu_1\right)}}{|\la_1-\mu_1|}\right),
\end{equation}
and in table \ref{stability1} we present the necessary conditions
for their existence.
\begin{table*}[t]
\begin{center}
\begin{tabular}{|c|c|c|c|c|c|c|c|}
\hline
 Cr. P.& $x_c$ & $y_c$ & Existence & Stable for & $\Omega_\s$ &  $w_{tot}$ & Acceleration   \\
\hline \hline
 A&  $x_{c1}$ & $y_{c1}$ & Always & $\la_1\left(\mu_1-\la_1\right)<3$ & 1 &
 $-\frac13\left(3+\la_1^2\right)$ & Always \\
B&  $x_{c2}$ & $y_{c2}$ & $\min\{\mu_1^2-3, \la_1^2+3\}\geq
\la_1\mu_1$, & Never & $\frac{\mu_1^2-\la_1
\mu_1-3}{(\la_1-\mu_1)^2}$ &
 $\frac{\mu_1}{\la_1-\mu_1}$ & $\mu_1<0,\, \mu_1<\la_1<-2\mu_1$ \\
 &    &   &  $\mu_1\neq\la_1$ &  &  &
 & $\mu_1>0,\, -2\mu_1<\la_1<\mu_1$\\
\hline
\end{tabular}
\end{center}
\caption[crit]{\label{stability1} The real and physically
meaningful critical points of Model 1 and their behavior.}
\end{table*}
 The $2\times2$ matrix ${\bf
{Q}}$ of the linearized perturbation equations writes:
\begin{widetext}
\[{\bf {Q}}= \left[
\begin{array}{cc}
 \frac{1}{2} \left(-9 x_c^2-2 \sqrt{6} \mu_1 x_c-3 \left(y_c^2+1\right)\right) & y_c \left(\sqrt{6} (\mu_1-\la_1)-3 x_c\right) \\
 -\frac{1}{2} y_c \left(6 x_c+\sqrt{6} \la_1\right) & \frac{1}{2} \left(-9 y_c^2-x_c \left(3 x_c+\sqrt{6} \la_1\right)+3\right)
\end{array}
 \right].\]
 \end{widetext}
Therefore, for each critical point of table \ref{stability1}, we
examine the signs of the real parts of the eigenvalues of ${\bf
{Q}}$, which determine the type and stability of this specific
critical point. In table \ref{stability1} we present the results
of the stability analysis. In addition, for each critical point we
calculate the values of $w_{tot}$ (given by relation
(\ref{wtot})), and of $\Omega_\phi$ (given by  (\ref{Omegas})).
Thus, knowing $w_{tot}$ we can express the acceleration condition
$w_{tot}<-1/3$ in terms of the model parameters.

The critical point A  exists always and it is either a saddle
point (the $Q$-eigenvalues have real parts of different sign) or
an attractor (the $Q$-eigenvalues have negative real parts). The
critical point B, if it exists, it is always a saddle point. The
cosmological  model at hand admits another critical point, namely
C, which is unphysical  since it leads to $\Omega_\phi<0$. This
point has coordinates $\left(x_{c3}=-\sqrt{\frac{2}{3}}\mu_1,\
y_{c3}=0\right)$ and it is either a saddle point or an attractor.
If $\mu_1(\mu_1-\lambda_1)>3/2$ it is an attractor and in this
case, although unphysical, it can attract an open set of orbits
from the interior of the physical region of the phase space.

\begin{figure}[ht]
\begin{center}
\hspace{0.4cm} \put(10,10){$C$} \hspace{0.4cm}
\put(55,85){$B$}\put(90,110){$A$}\put(230,15){$x$}\put(120,200){$y$}
\includegraphics[width=8cm,height=7cm]{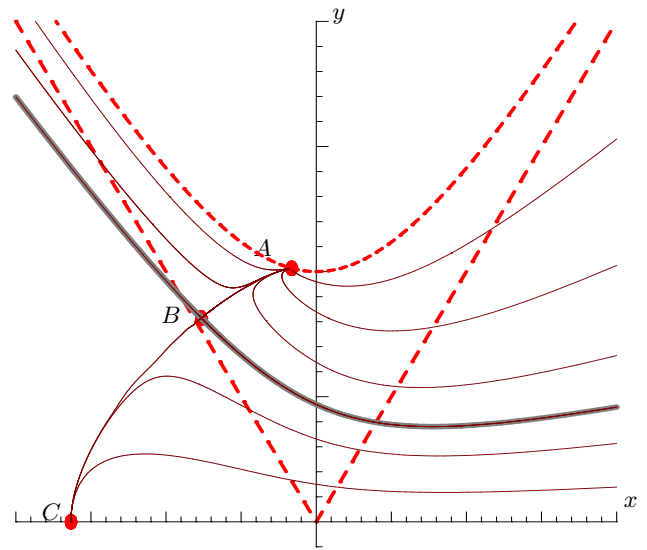}
\caption{(Color Online){\it Phase plane of Model 1 for the
parameter values $\la_1=0.4$ and $\mu_1=2$. The stable manifold of
B (thick curve) divides the physical part of the phase space
(region corresponding to $0\leq \Omega_\phi\leq1$, bounded by the
dashed (red) curves) in two regions. The orbits initially below
this curve converge towards $C.$ The orbits initially above this
curve converge to A.}} \label{Fig1}
\end{center}
\end{figure}

\begin{figure}[ht]
\begin{center}
\hspace{0.4cm} \put(90,15){$C$} \hspace{0.4cm}
\put(90,115){$A$}\put(230,15){$x$}\put(120,200){$y$}
\includegraphics[width=8cm,height=7cm]{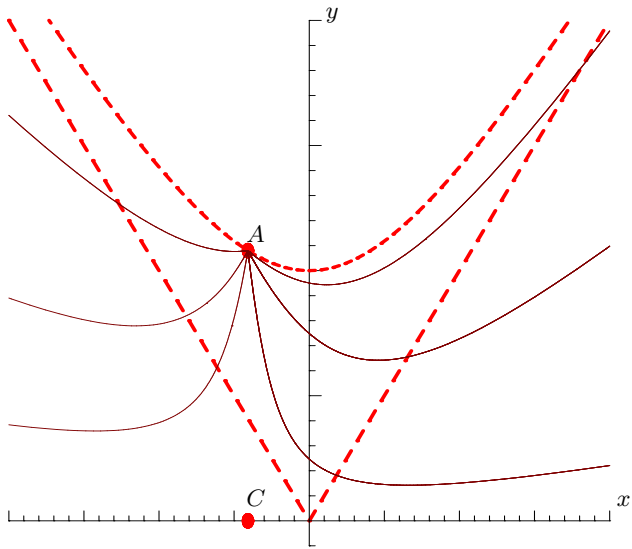}
\caption{(Color Online){\it Phase plane of Model 1 for the
parameter values $\la_1=1$ and $\mu_1=0.5$. In this case the
critical point B does not exist and all orbits initially at the
physical region converge to A. The dashed (red) curves bound the
physical part of the phase space, that is corresponding to
 $0\leq \Omega_\phi\leq1$.}} \label{Fig2}
\end{center}
\end{figure}

In order to present this behavior more transparently, we
 evolve the autonomous system (\ref{autonomous1}) numerically for
the parameters $\la_1=0.4$ and $\mu_1=2$, and the results are
shown in figure \ref{Fig1}.
Depending on which region of the phase-space does the system
initiates, it lies in the basin of attraction of either A or C,
and thus it is attracted by one or the other point. In particular,
the orbits initially below the stable manifold of B-points
converge towards C, while the orbits initially above this curve
converge to A. Interestingly, A is not the global attractor for
points at the physical region (region corresponding to $0\leq
\Omega_\phi\leq1$,  bounded by the dashed (red) curves). However,
if
$\frac{\la_1}{2}-\frac{\sqrt{6+\la_1^2}}{2}<\mu<\frac{\la_1}{2}+\frac{\sqrt{6+\la_1^2}}{2},$
point C is always a saddle one and B does not exist. Thus, in this
case A is the attractor for all the points located at the physical
region. This behavior is presented in figure \ref{Fig2}.
Finally, for completeness we mention that in the trivial case
$\mu_1=0$ the origin is also a saddle point. It represents
matter-dominated universe ($\Omega_{DM}\equiv
\frac{\kappa^2\rho_{DM}}{3 H^2}=1$) with $\phi$-independent dark
matter particle mass.

\subsection{Model 2: Power-law potential and power-law-dependent dark-matter particle mass}
\label{mod2a}

Inserting the auxiliary variables (\ref{auxilliary}) into the
equations of motion (\ref{rhodmeom}), (\ref{phiddot}), (\ref{FR1})
and (\ref{FR2}), we result in the following autonomous system:
\begin{eqnarray}
x'&=&-
3x+\frac{3}{2}x (1-x^2-y^2)-\frac{\lambda_2y^2 z}{2}-\frac{\mu_2}{2}z(1+x^2-y^2)\nonumber\\
y'&=&\frac{3}{2}y (1-x^2-y^2)-\frac{\lambda_2
xyz}{2}\nonumber\\
z'&=&-xz^2\label{autonomous2}.
\end{eqnarray}
The real and physically meaningful critical points are
\begin{eqnarray}
&&\left(x_{c4}=0,\ y_{c4}=0,\ z_{c4}=0\right), \nonumber\\
&&\left(x_{c5}=0,\ y_{c5}=1,\ z_{c5}=0\right),
\end{eqnarray}
and in table \ref{stability2} we present the necessary conditions
for their existence.
\begin{table*}[t]
\begin{center}
\begin{tabular}{|c|c|c|c|c|c|c|c|c|}
\hline
 Cr. P.& $x_c$ & $y_c$ & $z_c$ & Existence & Stable manifold & $\Omega_\s$ &  $w_{tot}$ & Acceleration   \\
\hline \hline
 D&  $x_{c4}$ & $y_{c4}$ &$z_{c4}$ & Always  &1-Dimensional   &  0
 &0
    & Never  \\
E&  $x_{c5}$ & $y_{c5}$ &$z_{c5}$ & Always  & 2-Dimensional  & 1 &
-1
& Always \\
\hline
\end{tabular}
\end{center}
\caption[crit]{\label{stability2} The real and physically
meaningful critical points of Model 2 and their behavior.}
\end{table*}
The $3\times3$ matrix ${\bf {Q}}$ of the linearized perturbation
equations writes:
\begin{widetext}
\[{\bf {Q}}= \left[
\begin{array}{ccc}
 \frac{1}{2} \left(-9 x_c^2-2 z_c \mu_2 x_c-3 y_c^2-3\right) & y_c (z_c (\mu_2-\la_2)-3 x_c) & \frac{1}{2}
   \left(y_c^2 (\mu_2-\la_2)-\left(x_c^2+1\right) \mu_2\right) \\
 -\frac{1}{2} y_c (6 x_c+z_c \la_2) & \frac{1}{2} \left(-9 y_c^2-x_c (3 x_c+z_c \la_2)+3\right) & -\frac{1}{2} x_c
 y_c
   \la_2 \\
 -z_c^2 & 0 & -2 x_c z_c
\end{array}
 \right].\]
 \end{widetext}
In this case, the critical points are non-hyperbolic, that is
there exists always at least a zero eigenvalue. We mention that
for non-hyperbolic critical points the result of linearization
cannot be applied in order to investigate the local stability of
the system (the system can be unstable to small perturbations on
the initial condition or to small perturbations on the parameters)
\cite{reza,arrowsmith,wiggins}. However, it is possible to get
information about the existence and the dimensionality of the
stable manifold by applying the center manifold theorem
\cite{arrowsmith}. Doing so we deduce that the dimensionality of
the local stable manifold is 1 and 2 for D and E respectively. In
particular, the stable manifold of D is tangent, at the critical
point, to the $x$-axis, while the stable manifold of E is tangent,
at the critical point, to the  $xy$-plane. The existence of an 1D
stable manifold for D, implies that the orbits asymptotic to D as
$t\rightarrow -\infty$ are contained in either an unstable or
center manifold (each one of dimensionality 1, that is a curve).
There are some exceptional orbits converging to D as $t\rightarrow
+\infty$, but these have a zero measure. On the other hand, the
fact that E has a 2D stable manifold implies that there exists a
non-zero-measure set of orbits that converges to E as
$t\rightarrow +\infty$. Finally, there are some exceptional orbits
contained in its center manifold that cannot be classified by
linearization. In summary, using more sophisticated tools such as
the Normal Forms theorem \cite{arrowsmith}, we indeed find that
the center manifold of E attracts an open set of orbits provided
$\la_2\leq 0$. On the other hand, if $\la_2> 0$ the orbits located
near the center manifold of E blow up in a finite time. Since this
point does not allow for a solution of the coincidence problem (it
always possesses $\Omega_\phi=1$) we do not present the
aforementioned procedure in detail.

\begin{figure}[ht]
\begin{center}
\hspace{0.4cm} \put(100,15){$D$} \hspace{0.4cm}
\put(100,110){$E$}\put(230,15){$x$}\put(120,200){$y$}
\includegraphics[width=8cm,height=7cm]{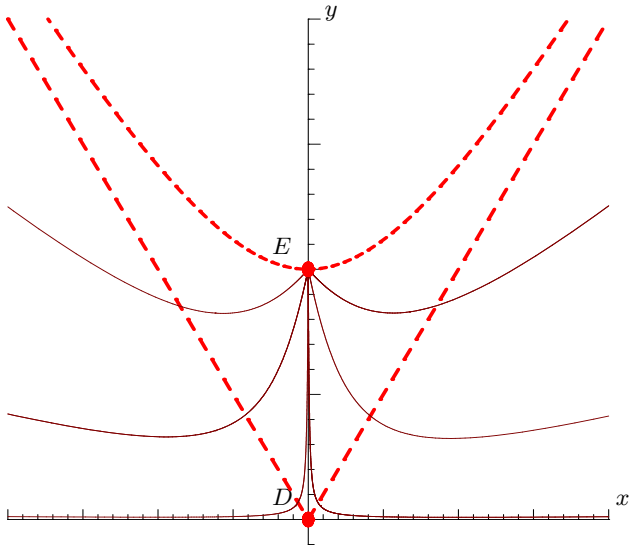}
\caption{(Color Online){\it xy-projection of the phase-space of
Model 2, for the parameter values $\la_2=-0.5$ and $\mu_2=0.5$.
The critical point E (representing de Sitter solutions) is the
attractor of the system. The dashed (red) curves bound the
physical part of the phase space, that is corresponding to
 $0\leq \Omega_\phi\leq1$.}} \label{Fig3}
\end{center}
\end{figure}

Numerical investigation reveals the above features.
 In fig.~\ref{Fig3} we depict orbits projected in the xy-plane,
as they arise from numerical evolution in the case of $\la_2=-0.5$
and $\mu_2=0.5$.

\subsection{Model 3: Power-law potential and exponentially-dependent dark-matter particle mass}

\begin{figure}[ht]
\begin{center}
\hspace{0.4cm} \put(20,15){$H$} \hspace{0.4cm}
\put(60,75){$G$}\put(105,105){$F$}\put(230,15){$x$}\put(120,200){$y$}
\includegraphics[width=8cm,height=7cm]{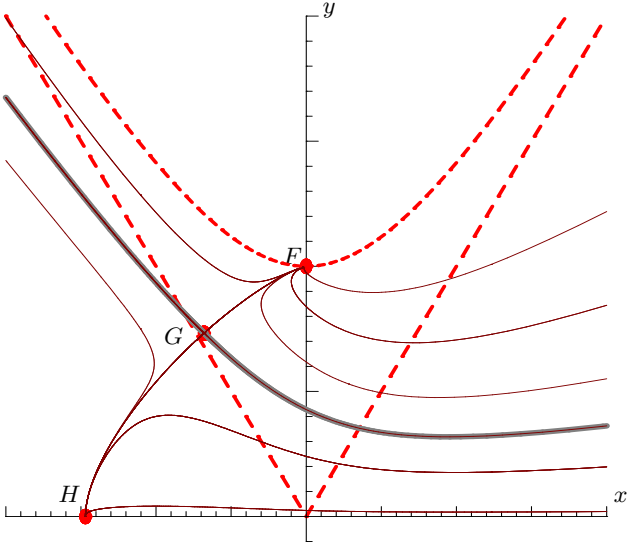}
\caption{(Color Online){\it xy-projection of the phase-space of
Model 3 for the parameter values $\la_2=1$ and $\mu_1=1.8$. The
stable manifold of G (thick curve) divides the physical part of
the phase space (region bounded by the dashed (red) curves) in two
regions. The orbits initially below this curve converge towards H,
while those initially above this curve converge towards F. }}
\label{Fig4}
\end{center}
\end{figure}

In this case the autonomous system reads:
\begin{eqnarray}
x'=-
3x+\frac{3}{2}x (1-x^2-y^2)-\frac{\lambda_2y^2 z}{2}-\ \ \ \ \ \ \   \ \ \ \ \nonumber\\-\sqrt{\frac{3}{2}}\mu_1(1+x^2-y^2)\nonumber\\
y'=\frac{3}{2}y (1-x^2-y^2)-\frac{\lambda_2 xyz}{2}\, \ \ \ \ \ \
\ \ \ \ \ \ \ \ \ \ \ \ \   \ \ \ \
\nonumber\\
z'=-xz^2. \ \ \ \ \   \ \ \  \ \ \ \   \ \ \ \ \ \ \ \ \ \ \ \ \ \
\ \ \ \   \ \ \ \ \ \ \ \ \ \ \   \ \ \ \ \ \ \ \
\label{autonomous3}
\end{eqnarray}
The real and physically meaningful critical points are
\begin{eqnarray}
&&\left(x_{c6}=0,\ y_{c6}=1,\ z_{c6}=0\right), \nonumber\\
&&\left(x_{c7}=-\frac{\sqrt{\frac{3}{2}}}{\mu_1},\
y_{c7}=\sqrt{1-\frac{3}{2\mu_1^2}},\ z_{c7}=0\right),
\end{eqnarray}
and the necessary conditions for their existence are shown in
table \ref{stability3}.
\begin{table*}[t]
\begin{center}
\begin{tabular}{|c|c|c|c|c|c|c|c|c|}
\hline
 Cr. P.& $x_c$ & $y_c$ & $z_c$ & Existence & Stable manifold & $\Omega_\s$ &  $w_{tot}$ & Acceleration   \\
\hline \hline
 F&  $x_{c6}$ & $y_{c6}$ &$z_{c6}$ & Always  &2-Dimensional   &  1
 &-1
    & Always  \\
G&  $x_{c7}$ & $y_{c7}$ &$z_{c7}$ & $|\mu_1|>\sqrt{3}$  &
1-Dimensional & $1-\frac{3}{\mu_1^2}$ & -1
& Always \\
\hline
\end{tabular}
\end{center}
\caption[crit]{\label{stability3} The real and physically
meaningful critical points of  Model 3 and their behavior.}
\end{table*}
The $3\times3$ matrix ${\bf {Q}}$ of the linearized perturbation
equations writes:
\begin{widetext}
\[{\bf {Q}}= \left[\begin{array}{ccc}
 \frac{1}{2} \left(-9 x_c^2-2 \sqrt{6} \mu_1 x_c-3 \left(y_c^2+1\right)\right) & y_c \left(-3 x_c-z_c \la_2+\sqrt{6}
   \mu_1\right) & -\frac{y_c^2 \la_2}{2} \\
 -\frac{1}{2} y_c (6 x_c+z_c \la_2) & \frac{1}{2} \left(-9 y_c^2-x_c (3 x_c+z_c \la_2)+3\right) & -\frac{1}{2} x_c
 y_c
   \la_2 \\
 -z_c^2 & 0 & -2 x_c z_c
\end{array}
 \right].\]
 \end{widetext}
In the model at hand, all critical points are non-hyperbolic and
the dimensionality of their stable manifold is presented in table
\ref{stability3}. Additionally, we mention that there exists also
an unphysical critical point H, with coordinates
$\left(x_{c8}=-\sqrt{\frac{2}{3}}\mu_1,\ y_{c8}=0\
z_{c8}=0\right)$. Its stable manifold is 2D if $|\mu_1|>
\sqrt{\frac{3}{2}}$, and 1D if $|\mu_1|< \sqrt{\frac{3}{2}}$. For
the choice $|\mu_1|> \sqrt{3},$ the orbits initially below the
stable manifold of G converge towards H. The orbits initially
above this curve converge towards F. This behavior is depicted in
fig. \ref{Fig4}, which has arisen from numerical evolution using
$\la_2=1$ and $\mu_1=1.8$.
If we restrict ourselves in the region $|\mu_1|<
\sqrt{\frac{3}{2}}$, then the critical point G does not exists and
thus there are not scaling solutions. In this case F is indeed the
attractor for a positive-measure set of initial conditions.
Moreover, there exist exceptional orbits contained on a 1D center
manifold of F whose dynamical behavior cannot be anticipated from
the linear analysis. However, since this scenario does not lead to
a solution of the coincidence problem ($\Omega_\phi=1$ always) we
do not present an advanced stability analysis for F.\\

\subsection{Model 4: Exponential potential and  power-law-dependent dark-matter particle mass}

\begin{figure}[ht]
\begin{center}
\hspace{0.4cm} \put(100,15){$I$} \hspace{0.4cm}
\put(90,120){$J$}\put(230,15){$x$}\put(120,200){$y$}
\includegraphics[width=8cm,height=7cm]{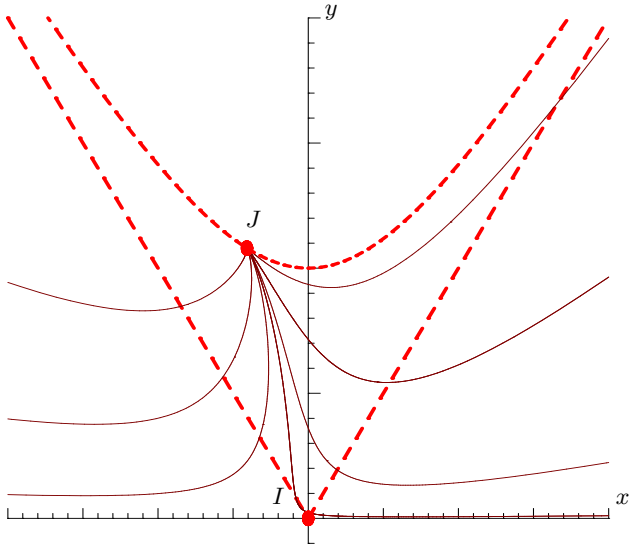}
\caption{(Color Online){\it xy-projection of the phase-space of
Model 4 for the parameter values $\la_1=1$ and $\mu_2=1.8$. The
critical point J (corresponding to a super-accelerating universe)
attracts all the orbits in this invariant set. The dashed (red)
curves bound the physical part of the phase space, that is
corresponding to
 $0\leq \Omega_\phi\leq1$.}} \label{Fig5}
\end{center}
\end{figure}

In this case the
 autonomous system writes:
\begin{eqnarray}
x'=-
3x+\frac{3}{2}x (1-x^2-y^2)-\sqrt{\frac{3}{2}}\la_1\, y^2-\ \ \  \ \ \ \ \nonumber\\-\frac{\mu_2}{2}z(1+x^2-y^2)\nonumber\\
y'=\frac{3}{2}y (1-x^2-y^2)- \sqrt{\frac{3}{2}}\la_1\, x y\ \ \ \
\ \ \ \
\ \ \ \ \ \ \ \ \ \ \  \ \nonumber\\
z'=-xz^2.    \ \ \ \ \ \ \ \ \ \ \   \ \ \ \ \ \ \ \ \ \ \ \ \ \ \
\ \ \   \ \ \ \ \ \ \ \ \ \ \   \ \   \ \ \ \ \ \ \
\label{autonomous4}
\end{eqnarray}
 The real and
physically meaningful critical points
 are
\begin{eqnarray}
&&\left(x_{c9}=0,\ y_{c9}=0,\ z_{c9}=0\right), \nonumber\\
&&\left(x_{c10}=-\frac{\la_1}{\sqrt{6}},\
y_{c10}=\sqrt{1+\frac{\la_1^2}{6}},\ z_{c10}=0\right),
\end{eqnarray}
and in table \ref{stability4} we present the necessary conditions
for their existence.
\begin{table*}[t]
\begin{center}
\begin{tabular}{|c|c|c|c|c|c|c|c|c|}
\hline
 Cr. P.& $x_c$ & $y_c$ & $z_c$ & Existence & Stable manifold & $\Omega_\s$ &  $w_{tot}$ & Acceleration   \\
\hline \hline
 I&  $x_{c9}$ & $y_{c9}$ &$z_{c9}$ & Always  &1-Dimensional   & 0
 &0
    & Never  \\
J&  $x_{c10}$ & $y_{c10}$ &$z_{c10}$ & Always  & 2-Dimensional & 1
& $-\frac13 (3+\la_1^2)$
& Always \\
\hline
\end{tabular}
\end{center}
\caption[crit]{\label{stability4} The real and physically
meaningful critical points of Model 4 and their behavior.}
\end{table*}
The $3\times3$ matrix ${\bf {Q}}$ of the linearized perturbation
equations reads:
\begin{widetext}
\[{\bf {Q}}= \left[
\begin{array}{ccc}
 \frac{1}{2} \left(-9 x_c^2-2 z_c \mu_2 x_c-3 y_c^2-3\right) & y_c \left(-3 x_c-\sqrt{6}
 \la_1
 +z_c \mu_2\right) & -\frac{1}{2} \left(x_c^2-y_c^2+1\right) \mu_2 \\
 -\frac{1}{2} y_c \left(6 x_c+\sqrt{6} \la_1\right) & \frac{1}{2} \left(-9 y_c^2-x_c
 \left(3 x_c+\sqrt{6} \la_1\right)+3\right) & 0 \\
 -z_c^2 & 0 & -2 x_c z_c
\end{array}
 \right].\]
\end{widetext}
The aforementioned critical points are non-hyperbolic since at
least one eigenvalue of ${\bf {Q}}$ is always zero. Linear
analysis in not conclusive in these cases, but information about
the dimensionality of the stable manifold can be obtained by
applying the center manifold theorem \cite{arrowsmith}. The
corresponding results are shown in table \ref{stability4}. Since
both I and J cannot solve the coincidence problem
($\Omega_\phi=1$), we do not present the aforementioned analysis
in detail. Finally, in order to acquire a more transparent picture
of the phase-space behavior, we evolve the system numerically for
$\la_1=1$ and $\mu_2=1.8$ and we depict the results in fig.
\ref{Fig5}.

\section{Cosmological implications and discussion}
\label{cosmimpl}

Having performed a complete phase-space analysis of several
varying dark-matter-mass models, we can discuss the corresponding
cosmological behavior. A general remark is that this behavior is
radically different from the corresponding quintessence scenarios
with the same potentials and mass-functions
\cite{Anderson:1997un,interacting2,Casas:1991ky,quirosvamps,Zhang:2005rg,Amendola:1999er,Amendola:2001rc,Comelli2003}.
Additionally, a common feature of almost all the phantom models
previously studied is the existence of attractors with
$w_{\phi}\leq-1$ in the whole phase-space \cite{phant,Guo:2004vg},
and thus, independently of the specific scenario and of the
imposed initial conditions, the universe always lies below the
phantom divide, as it is expected for phantom cosmology. This
global behavior is not always realized  in the case of
exponentially dependent dark-matter mass, and additional
constraints must be imposed.

Apart form acquiring acceleration, in this work we examine whether
the above constructed varying dark-matter-mass models can solve or
alleviate the coincidence problem. Thus, assuming as usual that
the present universe is already at a late-time attractor, we
calculate $\Omega_\phi$ in all stable fixed points, and if
$0<\Omega_\phi<1$ then the coincidence problem is solved since
$\Omega_\phi$ and $\Omega_{DM}$ will be of the same order of
magnitude as suggested by observations. On the contrary,
$\Omega_\phi=1$ corresponds to a universe completely dominated by
dark energy, while $\Omega_\phi=0$ (that is $\Omega_{DM}=1$ ) to
one completely dominated by dark matter, both in contrast with
observations.

Finally, we mention that as long as the interaction responsible
for the varying dark-matter particle mass is not too strong, the
standard cosmology can be always recovered. On the other hand,
since we assume that the universe is currently at an attractor,
its state is independent of the initial conditions. Thus, we can
switch on the interaction and consider as initial conditions the
end of the known epochs of standard Big Bang cosmology, in order
to avoid disastrous interference.

\subsection{Model 1}

In this model the critical point B is unstable, and therefore it
cannot be a late-time cosmological solution. The only relevant
critical point is A, which is a stable fixed point for
$\la_1\left(\mu_1-\la_1\right)<3$. As can be seen from table
\ref{stability1}, it corresponds to an accelerating universe with
$\Omega_\phi=1$, that is to complete dark-energy domination. Thus,
this specific cosmological solution cannot solve the coincidence
problem. Furthermore, the fact that $w_{tot}$ is not only less
than $-1/3$, as required by the acceleration condition, but it is
always less than $-1$, leads to $\dot{H}>0$ at all times.
Therefore, this solution corresponds to a super-accelerating
universe \cite{Das:2005yj}, that is with a permanently increasing
$H$, resulting to a Big Rip. This behavior is common in phantom
cosmology \cite{phant,Briscese:2006xu}.

A remarkable feature of this model, as well as  of Model 3, is
that if there exist scaling solutions, then, for a wide region of
the parameter space, the stable manifold of the corresponding
critical point marks the basin of attraction of either a phantom
attractor or an unphysical attracting state. Thus, there exist an
open set of orbits of the physical region that converge to an
unphysical state instead to a phantom solution. This behavior was
revealed analytically and it was confirmed through numerical
elaboration, and seems to be typical in the case of
exponentially-dependent dark-matter mass in the phantom framework.
To avoid dealing with unphysical states, we can either restrict
the physical portion of the phase-space to the region above the
stable manifold of the scaling solutions, or restrict the
parameter-space itself. In both cases we obtain an additional
constraint, that was not present in previous studies of phantom
cosmology \cite{Guo:2004vg,Guo:2004xx,ChenSaridakis}, which
further weakens the applicability of the model.

In summary, Model 1, that is an exponential potential and an
exponentially-dependent dark-matter particle mass, cannot act as a
candidate for solving the coincidence problem.

\subsection{Model 2}

In  this case, both real and physically meaningful critical
points, namely D and E, have a stable manifold of smaller
dimensionality than that of the phase-space, and as was mentioned
in subsection \ref{mod2a} they have  very small probability to be
the late-time attractors of the system. However, even if the
cosmological evolution is managed to be attracted by these
solutions, the coincidence problem will not be solved, since D
represents a flat, non-accelerating universe dominated by dark
matter, and E correspond to de Sitter universe completely
dominated by dark energy. These critical points are located in the
region where the scalar field and the Hubble parameter diverge.
Divergencies in a cosmological scenario are represented as
asymptotic states, in particular associated with the past and
future asymptotic dynamics \cite{Wetterich:1994bg2,pastdynamics}.
In the present Model 2, due to the non-compactness of the
phase-space, such a behavior can lead either to an asymptotic
state acquired at infinite time, or to a singularity reached at a
finite time. If $H\rightarrow\infty$ or
$\rho_\phi\rightarrow\infty$ at $t\rightarrow\infty$ then we
acquire an eternally expanding universe, while if
$H\rightarrow\infty$ at $t\rightarrow t_{BR}<\infty$ then the
universe results to a Big Rip \cite{Capozziello:2009hc}.

Therefore, power-law potentials with power-law-dependent
dark-matter particle masses, cannot solve the coincidence problem.

\subsection{Model 3}

In this model we see that the critical point F exists always,
while G exists only for $|\mu_1|>\sqrt{3}$. However, in both cases
the stable manifold is of smaller dimensionality than that of the
phase-space. Furthermore, in order to avoid the treatment of
unphysical attracting states we have to impose the additional
constraint
 $|\mu_1|< \sqrt{\frac{3}{2}}$. For this choice
of parameters, G does not exists and thus there are not scaling
solutions, while F is the attractor for a positive-measure set of
initial conditions. Point F corresponds to a dark-energy dominated
de Sitter universe, while G to a flat accelerating universe with
$\Omega_\phi=1-\frac{3}{\mu_1^2}$, that is with $0<\Omega_\phi<1$
in the region that it exists. In both points the phantom field
diverges. However, even if G possesses $0<\Omega_\phi<1$, it can
not solve the coincidence problem since it is not a relevant
late-time attractor.

In summary, power-law potentials with exponentially-dependent
dark-matter particle masses cannot solve or even alleviate the
coincidence problem.

\subsection{Model 4}

In this case, the critical points I and J exist always. The point
I corresponds to a flat, non-accelerating, matter-dominated
universe. J corresponds to a dark-energy dominated universe, that
super-accelerates \cite{Das:2005yj}. However, similarly to the
previous cases, the stable manifolds of I and J are 1D or 2D
respectively, and thus almost all orbits of the cosmological
system cannot be attracted by them  at late times. In addition,
they cannot lead to $0<\Omega_\phi<1$. Therefore, an exponential
potential and a power-law-dependent dark-matter particle mass,
cannot solve the coincidence problem.

\section{Conclusions}
\label{conclusions}

In the present work we investigated the phantom cosmological
scenario, with varying-mass dark-matter particles due to the
interaction between dark-matter and dark-energy sectors. In
particular, we performed a detailed phase-space analysis of
various models, with either exponentially or power-law dependent
dark-matter particle mass, in exponential or power-law scalar
field potentials. These functions cover a wide range of the
possible forms, and they correspond to the cases that can accept a
reasonable theoretical justification
\cite{Amendola:1999er,Comelli2003,Zhang:2005rg,Kneller:2003xg,Amendola:2001rc,Zhang:2005rg}.
In each case we extracted the critical points, we determined their
stability, and we calculated the basic cosmological observables,
namely the total equation-of-state parameter $w_{tot}$ and
$\Omega_{{\text{DE}}}$ (attributed to the phantom field). Our
basic goal was to examine whether there exist late-time
attractors, corresponding to accelerating universe and possessing
$\Omega_{{\text{DE}}}/\Omega_{{\text{DM}}}\approx{\cal{O}}(1)$,
thus satisfying the basic observational requirements.

Apart from the case of an exponential potential with an
exponentially-dependent dark-matter particle mass, which possesses
a relevant late-time (phantom) attractor, in all the other models
we found that physical, well-motivated solutions have a very small
chance to attract the universe at late times. In addition, in all
the examined cases, solutions having
$\Omega_{{\text{DE}}}/\Omega_{{\text{DM}}}\approx{\cal{O}}(1)$ are
not relevant attractors at late times. Therefore, summarizing, the
coincidence problem cannot be solved or even alleviated in
varying-mass dark matter particles models in the framework of
phantom cosmology, in a radical contrast with the corresponding
quintessence case \cite{Amendola:1999er,Comelli2003,Zhang:2005rg}.
This conclusion agrees with that of \cite{ChenSaridakis}, that
interacting phantom cosmology cannot solve the coincidence
problem. It seems that interacting phantom cosmology, either
directly or through the dependence of the dark-matter particle
mass, cannot fulfill the basic requirements that led to its
construction, that is to provide stable accelerating late-time
solutions which can solve the coincidence problem. An alternative
direction could be to consider a specially constructed potential
or dark-matter particle mass in order to solve the coincidence
problem, but this would imply significant loss of simplicity,
generality, and theoretical justification of the model.

The aforementioned conclusion has been extracted by the
negative-kinetic-energy realization of phantom, which does not
cover the whole class of phantom models. However, since it is a
qualitative statement it should intuitively be robust for general
phantom scenarios, too. Therefore, phantom cosmology with
varying-mass dark matter particles cannot easily act as a
successful candidate to describe dark energy.

\vskip .1in \noindent {\bf {Acknowledgments}}

 G. L wishes to thank
the MES of Cuba for partial financial support of this
investigation. His research was also supported by Programa
Nacional de Ciencias B\'asicas (PNCB).

\end{document}